\documentclass[12pt,a4paper]{article}

\usepackage{graphicx} % Required for inserting images
\usepackage{hyperref}
\usepackage{graphicx}
\usepackage{amssymb}
\usepackage{amsmath}
\usepackage{color}
\usepackage{url}

\title{\large\bf Non-Supersymmetric Baryogenesis from $U(1)$-Breaking Scalar Dynamics}

\author{\normalsize\bf Surendra Kumar Gour\thanks{\tt s.gour@iitg.ac.in} \ and Malay K. Nandy\thanks{\tt mknandy@iitg.ac.in \rm (Corresponding Author)} \\
 {\small\em Department of Physics, Indian Institute of Technology Guwahati, Guwahati 781 039, India}}
 
\date{\small (1 May 2026)} 

\begin{document}

\maketitle

\begin{abstract}
We present a non-supersymmetric mechanism for baryogenesis \linebreak driven by the nonlinear dynamics of a complex scalar field with generalized self-interaction potentials that explicitly break the global $U(1)$ symmetry. Specifically, three representative forms of the interaction potential are considered, which give rise to intricate nonlinear source terms in the evolution of the field components. In all cases, we show that these nonlinear source terms dynamically generate a nonzero Noether charge density from symmetric initial conditions, providing a purely dynamical origin of charge asymmetry. At late times, the charge density scales as $\sim t^{-3/2}$, leading to a constant baryon-to-photon ratio through dynamical freeze-in. While the qualitative behavior is robust across models, the quantitative features depend sensitively on the interaction structure. We find that one class of potentials yields a viable parameter space over a wide range of scalar masses, whereas another requires unrealistically suppressed mass scales. A third scenario stands out in that the final asymmetry is independent of the scalar mass and depends only on the coupling parameter, enhancing predictivity and allowing compatibility across a broad range of energy scales. Assuming efficient transfer of the generated asymmetry to the Standard Model sector and negligible washout effects, the mechanism can account for the observed baryon asymmetry.
\end{abstract}

\tableofcontents

%\begin{multicols}{2}

\section{Introduction}

Thermal history of the Universe after big bang nucleosynthesis (BBN) is well-understood and strongly supported by observations, forming a key part of the standard cosmological model. This framework faces several fundamental challenges, including the horizon, flatness, and monopole problems. The idea of inflation \cite{Starobinsky-1980, Guth-1981, Linde-1982, Albrecht-1982, Linde-1983} provides an elegant solution to these issues. However, inflation dilutes any preexisting asymmetry, thus does not offer an explanation for the observed baryonic abundance in the Universe. It is feasible that baryon-antibaryon asymmetry must generate after the inflationary epoch and it requires physical processes beyond those considered in standard theoretical frameworks. 

The phenomenon of baryogenesis seeks to require a theoretical framework that explains the origin of the baryon-antibaryon asymmetry observed in the Universe. This imbalance is quantified by the baryon-to-photon ratio,
\begin{equation}
    \eta_{B} = \frac{n_{b}-n_{\bar{b}}}{n_{\gamma}},
\end{equation}
where the quantities $n_{b}$, $n_{\bar{b}}$, and $n_{\gamma}$ correspond to the number densities of baryons, antibaryons, and photons, respectively. 

Observational data constrain the baryon-to-photon ratio $\eta_{B}$ in the range  $\sim~(5.5\text{--}6.8)\times 10^{-10}$, with a central estimate of $6.1\times 10^{-10}$. This is measured by BBN prediction and CMB WMAP observation \cite{Burles-PRD-2001, Burles-2001, Bennett-2003}, consistent with the Planck measurement \cite{Planck-2018, Planck-2024, PDG-2024} of the baryon density, $\Omega_bh^2 = 0.0224 \pm 0.0001$.

To explain the observed imbalance between baryon and antibaryon in the Universe, Sakharov \cite{Sakharov-1967} proposed a set of three fundamental conditions: baryon number violation, charge and charge-parity violation, and departure from thermodynamic equilibrium. Realization of these conditions can dynamically generate a baryon asymmetry in the early Universe. Over the years, a variety of theoretical mechanisms have been proposed to realize these conditions in concrete physical models. One of the earliest approaches is GUT baryogenesis \cite{Georgi-Glashow-1974,  Dimopoulos-1978, Yoshimura-1978-A, Yoshimura-1978-B, Weinberg-1979, Kolb-1980, Kolb-1990-book, Weinberg-book}, where the decay of superheavy particles in the early Universe generates a baryon asymmetry through CP-violating interactions. On the other hand, electroweak baryogenesis \cite{Kuzmin-1985, Shaposhnikov-1986, Cline-2006} operates at the scale of electroweak phase transition and involves non-perturbative processes such as the sphaleron process. Furthermore, the Affleck-Dine mechanism \cite{Affleck-Dine-1985, Allahverdi-2012} operates in supersymmetric settings, where scalar fields with baryon number evolve along flat directions of the potential. These fields develop large vacuum expectation values and subsequently decay into Standard Model particles, producing the observed baryonic asymmetry.

Recent studies have explored inflationary baryogenesis \cite{Raghavan-2001}, where the inflaton field itself is responsible for generating the matter-antimatter asymmetry, extending ideas related to the Affleck-Dine mechanism into the inflationary epoch. In particular, nonlinear effects at the end of inflation have been investigated as a source of inflaton-induced asymmetry \cite{Lozanov-2014}. In these models \cite{Hertzberg-2014A, Hertzberg-2014B, Takeda-2015}, the inflaton is treated as a complex scalar field with a weakly broken global symmetry, allowing for CP violation through a dynamically evolving phase during inflation and reheating. The resulting asymmetry is subsequently transferred to the baryonic sector through inflaton decay, providing a viable explanation for the observed baryon asymmetry of the Universe.

In recent works, non-supersymmetric scenarios have been proposed in which the decay of two nearly degenerate superheavy scalar fields drives baryogenesis \cite{Babichev-2018, Babichev-2019}. Coupled to the inflaton, these fields develop nonzero vacuum expectation values during inflation. A weakly broken global $U(1)$ symmetry produces a small mass splitting between them, resulting in the dynamical emergence of a net baryon asymmetry after inflation.

An alternative minimal approach \cite{Lloyd-2021, Lloyd-2023} attributes the origin of the baryon asymmetry to a complex inflaton condensate. Here, baryon-violating quadratic terms in the inflaton potential, combined with the Affleck-Dine mechanism, generate a dynamical asymmetry. At late times, these terms produce oscillations in the baryon asymmetry of the condensate. The final asymmetry transferred during reheating is reduced to the observed magnitude through averaging over these oscillations.

In a recent work \cite{Gour-2026}, a non-supersymmetric mechanism for baryogenesis was explored based on the nonlinear dynamics of a complex scalar field with a $U(1)$-breaking self-interaction potential. This setup naturally leads to dynamical generation of an asymmetry that stabilizes at late times across a range of mass scales. 

In this work, we consider a more general class of interactions for the $U(1)$-breaking complex scalar field. In particular, we analyze three distinct forms of interaction and study their impact on the generation of the asymmetry. This leads to more intricate nonlinear source terms in the dynamical equations. Notably, for one class of interaction, the final baryon-to-photon ratio becomes independent of the scalar mass, revealing a striking and nontrivial feature of the framework.

The remainder of this paper is organized as follows. Section~\ref{sec-model} presents the generalized scalar field models. In Section~\ref{sec-noether}, we derive the dynamical evolution equations governing the Noether charge and baryon asymmetry. Section~\ref{sec-baryon} is devoted to the formulation of the baryon-to-photon ratio. The numerical analysis and corresponding results are discussed in Section~\ref{sec-numerical}. Finally, Section~\ref{sec-disc} provides a discussion and concluding remarks.

\section{Scalar Field Model}
\label{sec-model}

For the analysis of baryogenesis, we consider a complex scalar field $\Phi$ whose action is given by 
\begin{equation}
S = \int d^{4} x \sqrt{-g} \left[-\frac{1}{2}\, g^{\mu\nu}\partial_{\mu}\Phi^{*}\partial_{\nu}\Phi - \frac{1} {2}\, M^2 |\Phi|^2 -V(\Phi,\Phi^{*}) \right],
\label{eq-action}
\end{equation}
where $M$ is the mass of the scalar field. We model the self-interaction potential $V(\Phi,\Phi^{*})$  as 
\begin{equation}
   	V(\Phi, \Phi^{*})=   \frac{\lambda}{4!}  \frac{\left(\Phi\Phi^{*}\right)^{n}}{M_{P}^{2n}} \left[\Phi^{3}+ \Phi^{*3}\right] \left(\Phi + \Phi^{*}\right),
\label{eq-potential}
\end{equation}
where $\lambda$ is the strength of the potential, $M_{P}$ is the reduced Planck mass, and $n$ is a positive integer. This choice for the potential is inspired by the baryogenesis model of Dimopoulos and Susskind \cite{Dimopoulos-1978}.

The non-interacting part of the action \ref{eq-action} admits a $U(1)$ symmetry which implies invariance of the Noether charge. Since the potential $V(\Phi, \Phi^{*})$ breaks the $U(1)$ symmetry, it  generates an asymmetry in the Noether charge.

Evidence from the Cosmic Microwave Background (CMB) radiation shows that the Universe is homogeneous and isotropic on large scales. Consequently, we employ the line element of a flat FLRW Universe \cite{Friedmann-1922, Lema-1927, Robertson-1936, Walker-1937, Barrow-2012}, which is given by  
\begin{equation}
			ds^2 = -dt^2 + a^2 (dx^2 + dy^2 + dz^2),
\end{equation}
where $a(t)$ is the scale factor and $t$ is the cosmic time.

Furthermore, homogeneity and isotropy indicate that the scalar field is solely time-dependent. Consequently, the total action reduces to
\begin{equation}
S = V \int dt\, a^3 \left[\frac{1}{2}\,|\dot\Phi|^2  - \frac{1} {2}\, M^2 |\Phi|^2 - \frac{\lambda}{4!} \frac{\left(\Phi\Phi^{*}\right)^{n}}{M_{P}^{2n}}\Bigl(\Phi^3+\Phi^{*3}\Bigr)\Bigl(\Phi+\Phi^*\Bigr) \right],
\label{eq-action2}
\end{equation}
where $V$ is a fiducial volume, and an overdot indicates  differentiation with respect to time $t$.

Writing the complex field as $\Phi= u+iv$, the action \ref{eq-action2} gives the Lagrangian
\begin{equation}
     L= Va^3 \left[\frac{1}{2}(\dot{u}^2+ \dot{v}^2) -\frac{1}{2}M^{2}\left(u^{2}+v^{2}\right) -\frac{\lambda}{6 M_{P}^{2n}}\, \left(u^{2}+v^{2}\right)^{n}\left(u^{4}-3u^{2}v^{2}\right)\right],
\end{equation}
so that we obtain the equations of motion in the forms 
\begin{equation}
     \ddot{u} + 3 H\dot{u} + M^{2} u = \frac{\lambda}{6\,M_{P}^{2n}} \left[ \left(6uv^{2}-4u^{3}\right) \left(u^{2}+v^{2}\right)^{n} +2nu \left(3u^{2}v^{2}-u^{4}\right) \left(u^{2}+v^{2}\right)^{n-1}\right]
 \label{eq-u}     
\end{equation}
and
\begin{equation}
     \ddot{v} + 3 H\dot{v} + M^{2} v =\frac{\lambda}{6\,M_{P}^{2n}}\left[ 6u^{2}v\left(u^{2}+v^{2}\right)^{n}  +2nv \left(3u^{2}v^{2}-u^{4}\right)\left(u^{2}+v^{2}\right)^{n-1}\right],
  \label{eq-v}    
\end{equation}
upon employing the Euler-Lagrange equation for the real-valued scalars  $u$ and $v$.

In addition to their coupled nature, the governing equations \ref{eq-u} and \ref{eq-v} exhibit significant nonlinearity. Therefore, analytical solutions are impractical, and we shall employ numerical integration of these equations. Moreover, for different values of $n$, the form of the coupled equations will vary, and distinct transformations are required to render each case dimensionless.

For numerical implementation, we reformulate the equations in terms of dimensionless variables, defined by
\begin{equation}
x_{n} = \left(\frac{\lambda}{M^{2}M_{P}^{2n}}\right)^{\frac{1}{2n+2}}u, \qquad
y_{n} = \left(\frac{\lambda}{M^{2}M_{P}^{2n}}\right)^{\frac{1}{2n+2}} v , \quad \text{and} \quad \tau =  Mt.
\label{eq-xytau}
\end{equation}
As a result, the dynamical equations \ref{eq-u} and \ref{eq-v} can be written as
\begin{equation}
\frac{d^2x_{n}}{d\tau^2}+\frac{3}{2\tau}\frac{dx_{n}}{d\tau}+x_{n} = \frac{1}{6}\left[(6x_{n}y_{n}^2 - 4x_{n}^3)(x_{n}^2+y_{n}^2)^n + 2n x_{n}(3x_{n}^2y_{n}^2 - x_{n}^4)(x_{n}^2+y_{n}^2)^{n-1} \right]
\label{eq-x}
\end{equation}
and 
\begin{equation}
\frac{d^2y_{n}}{d\tau^2}+\frac{3}{2\tau}\frac{dy_{n}}{d\tau}+y_{n}=\frac{1}{6}\left[6x_{n}^2 y_{n} (x_{n}^2+y_{n}^2)^n + 2n y_{n}(3x_{n}^2y_{n}^2 - x_{n}^4)(x_{n}^2+y_{n}^2)^{n-1} \right],
\label{eq-y}
\end{equation}
where we have assumed a radiation-dominated background with Hubble parameter $H=\frac{1}{2t}$. 

It is evident from the Eqs.~\ref{eq-x} and \ref{eq-y} that their right-hand sides act  like source terms, which do not possess symmetry under the exchange $x_{n}\leftrightarrow y_{n}$. Consequently, the components $x_n(\tau)$ and $y_n(\tau)$ evolve differently, leading to an imbalance between the two field-components.

\section{Evolution of Noether Charge and Baryon Asymmetry}
\label{sec-noether}

The Noether charge represents a conserved quantity arising from the $U(1)$ symmetry of the non-interacting sector of the action \ref{eq-action}, from which the Noether charge density $\rho$ is obtained as
\begin{equation}
    \rho = \frac{i}{2}\left(\Phi\,\dot{\Phi}^* - \Phi^*\,\dot{\Phi}\right).
\label{eq-rho}    
\end{equation}

This $U(1)$ symmetry is explicitly broken due to the self-interaction potential \ref{eq-potential}, leading to a time-evolution of the Noether charge density $\rho$. This time-evolution can be captured from the equations of motion for $u$ and $v$, given by \ref{eq-u} and \ref{eq-v}. 

Expressing the scalar field as $\Phi = u + i v$, the Noether charge density \ref{eq-rho} reduces to
\begin{equation}
\rho = u\,\dot{v} - v\,\dot{u},
\label{eq-rho1}
\end{equation}
which directly yields
\begin{equation}
\dot{\rho} = u\,\ddot{v} - v\,\ddot{u},
\label{eq-dotrho}
\end{equation}
thereby governing the time evolution of $\rho(t)$.

Using equations of motion \ref{eq-u} and \ref{eq-v} in  \ref{eq-dotrho} leads to
\begin{equation}
    \dot{\rho}_{n} + \frac{3}{2t}\rho_{n} = \frac{\lambda}{M_{P}^{2n}}\left(u^{2}+v^{2}\right)^{n} \left(\frac{5}{3}u^{3}v-uv^{3}\right),
    \label{eq-dotrho1}
\end{equation}
which describes the evolution of the Noether charge density in the presence of interactions.

Performing a transformation to the dimensionless variables defined in Eq.~\ref{eq-xytau}, the dimensionless counterpart of Noether charge density \ref{eq-rho1} 
is defined as 
\begin{equation}
\tilde{\rho}_{n}(\tau)= x_{n}\frac{dy_{n}}{d\tau} - y_{n}\frac{dx_{n}}{d\tau},
\label{eq-rhon1}
\end{equation}
where
\begin{equation}
\tilde{\rho}_{n}(\tau)=\frac{1}{M}\left(\frac{\lambda}{M^{2}M_{P}^{2n}}\right)^{\frac{1}{n+1}}\rho_{n}(\tau),
\label{eq-rhon}
\end{equation}
and its time-evolution \ref{eq-dotrho} translates to 
\begin{equation} 
\frac{d\tilde{\rho}_{n}}{d\tau}= x_{n}\frac{d^2y_{n}}{d\tau^2} - y_{n}\frac{d^2x_{n}}{d\tau^2}.
\label{eq-til-rho}
\end{equation}

\begin{figure}
\centering
\includegraphics[width=0.70\textwidth]{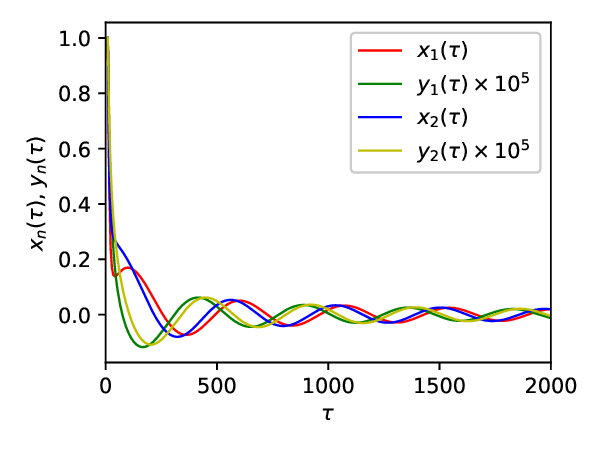}
  \caption{Temporal evolution of the dimensionless real and imaginary components, $x_{n}(\tau)$ and $y_{n}(\tau)$, of the complex scalar field, obtained from numerical solution of the coupled differential equations \ref{eq-x} and \ref{eq-y} for $n=1,2$.}
\label{fig-xy-12} 
\end{figure}

Using \ref{eq-x} and \ref{eq-y}, Eq.~\ref{eq-til-rho} takes the form 
\begin{equation}
    \frac{d\tilde{\rho}_{n}}{d\tau}+\frac{3}{2\tau}\tilde{\rho}_{n} =\left(x_{n}^{2}+y_{n}^{2}\right)^{n} \left(\frac{5}{3}x_{n}^{3}y_{n}-x_{n}y_{n}^{3}\right),
    \label{eq-til-rho1}
\end{equation}
where the non-linear term on the right-hand side acts as a source which is responsible for dynamical evolution of the dimensionless Noether charge density $\tilde{\rho}_{n}(\tau)$ despite the Hubble expansion term on the left-hand side. This can be confirmed by simultaneous numerical integration of the coupled differential equations \ref{eq-x} and \ref{eq-y} to obtain dynamical evolution of $x_n(\tau)$ and $y_n(\tau)$ so that the function $\tilde{\rho}_{n}(\tau)$ can be calculated from numerical integration of Eq.~\ref{eq-til-rho1} supplemented with Eq.~\ref{eq-rhon1}.
 
A concrete realization of the transfer of the generated Noether charge asymmetry to the baryonic sector can be consistently achieved through effective baryon-number-violating interactions, coupling the complex scalar field $\Phi$ to the Standard Model degrees of freedom. During the post-inflationary reheating epoch, these interactions can efficiently convert the stored $U(1)$ asymmetry into a net baryon asymmetry in the Standard Model sector. Provided that washout effects are sufficiently suppressed, the resulting baryon number becomes frozen in, yielding a conserved baryon-to-photon ratio at late times. This establishes a dynamical link between the scalar-sector asymmetry and the observed baryon asymmetry of the Universe.

\begin{figure}
\centering
\includegraphics[width=0.70\textwidth]{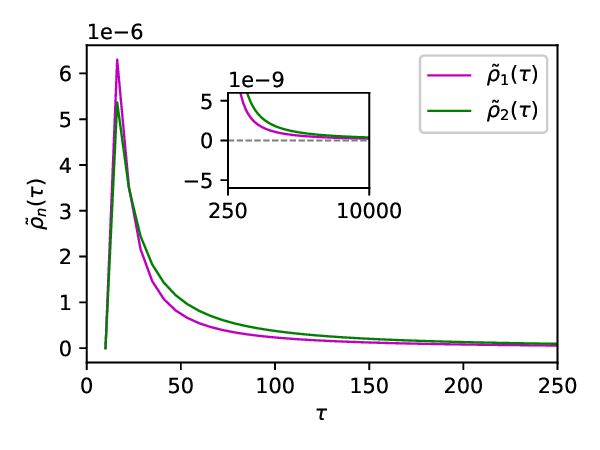}
  \caption{Temporal evolution of the dimensionless Noether charge density $\tilde{\rho}_{n}(\tau)$ for $n=1,2$, obtained from simultaneous numerical solution of Eq.~\ref{eq-til-rho1}, \ref{eq-x}, and \ref{eq-y}, supplemented with Eq.~\ref{eq-rhon1}. The inset shows the long-time behaviour.}
  \label{fig-r-til-12}
\end{figure}

\section{Baryon-to-Photon ratio}
\label{sec-baryon}

In order to obtain the baryon-to-photon ratio, time evolution of the photon number density $n_\gamma$ in the radiation-dominated (RD) epoch is required. This is related to the  temperature $T$ as 
\begin{equation}
n_{\gamma} = \frac{2\zeta(3)}{\pi^2}T^3,
\label{eq-ngamma}
\end{equation}
where $ \zeta(3) \simeq 1.2021 $. In this RD epoch, the Friedmann equation is effectively given by
\begin{equation}
H^2 = \frac{8\pi G}{3}\rho_\gamma,
\label{eq-fried}
\end{equation}
where the radiation energy density $\rho_\gamma$ follows the Stefan-Boltzmann law,
\begin{equation}
\rho_\gamma = \frac{\pi^2}{30}g_*T^4,
\label{eq-rad}
\end{equation}
with \( g_* \) denoting the number of relativistic degrees of freedom.

\begin{figure}
\centering
\includegraphics[width=0.70\textwidth]{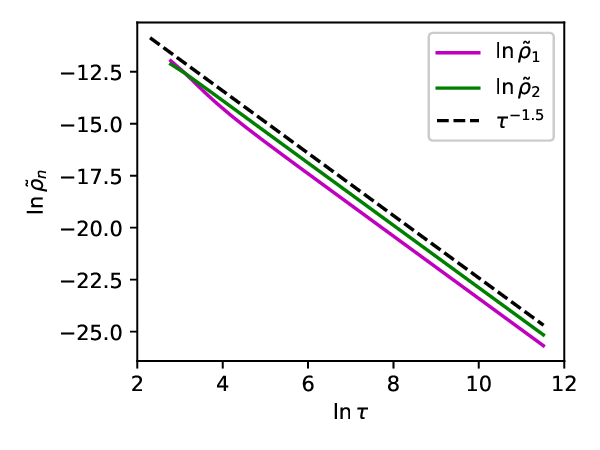}
  \caption{Logarithmic plots for the dimensionless Noether charge density, $\ln\tilde{\rho}_{n}(\tau)$ versus $\ln\tau$, for $n=1,2$. The dashed reference line has a slope of $-\frac{3}{2}$.}
  \label{fig-log-r-til-12}
\end{figure}

In the radiation-dominated era, the scale factor evolves as \( a(t) \propto t^{1/2} \), implying a Hubble parameter \( H(t) = \dot{a}/a = 1/(2t) \). Substituting this into Eq. \ref{eq-fried}, together with Eq. \ref{eq-rad}, yields the time-dependent temperature
\begin{equation}
T(t) = \left(\frac{45}{2\pi^2 g_*}\right)^{\frac{1}{4}}\left(\frac{M_P}{t}\right)^{\frac{1}{2}},
\end{equation}
where \( M_P = (8\pi G)^{-1/2} \) is the reduced Planck mass. Substituting this result into Eq.~\ref{eq-ngamma}, time-evolution of the photon number density turns out to be 
\begin{equation}
n_\gamma(t) = \frac{2\zeta(3)}{\pi^2}\left(\frac{45}{2\pi^2 g_*}\right)^{\frac{3}{4}}\left(\frac{M_P}{t}\right)^{\frac{3}{2}}.
\end{equation}

Introducing the dimensionless time variable  \( \tau = Mt \), the expression above can be rewritten as
\begin{equation}
n_\gamma(\tau) = \frac{2\zeta(3)}{\pi^{7/2}}\left(\frac{45}{2g_*}\right)^{3/4}(M_P M)^{3/2}\,\tilde{n}_\gamma(\tau),
\label{eq-photon}
\end{equation}
where the dimensionless photon number density is 
\begin{equation}
\tilde{n}_\gamma(\tau) = \tau^{-3/2}.
\label{eq-ngam}
\end{equation}

From Eqs.~\ref{eq-rhon} and \ref{eq-photon}, the baryon-to-photon ratio turns out to be 
\begin{equation}
\label{eq-eta}
\eta_{n}(t) = \frac{\rho_{n}(t)}{n_\gamma(t)}
= \frac{1}{\lambda^{\frac{1}{n+1}}}\frac{\pi^{7/2}}{2\zeta(3)}\left(\frac{2g_*}{45}\right)^{3/4} \frac{1}{M^{1/2} M_P^{3/2}}\left(M^2 M^{2n}_{P}\right)^{\frac{1}{n+1}}\,\tilde{\eta}_{n}(\tau),
\end{equation}
where 
\begin{equation}
\label{eq-til-eta}
\tilde{\eta}_{n}(\tau) = \frac{\tilde{\rho}_{n}(\tau)}{\tilde{n}_\gamma(\tau)}.
\end{equation}

\begin{figure}
\centering
\includegraphics[width=0.70\textwidth]{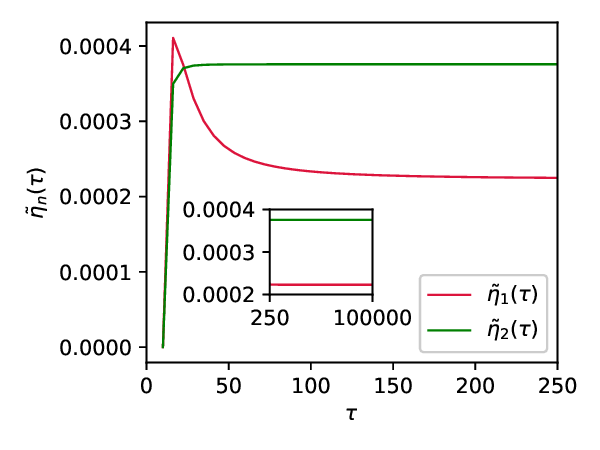}
\caption{Temporal evolution of $\tilde{\eta}_{n}(\tau) = \frac{\tilde{\rho}_{n}(\tau)}{\tilde n_\gamma(\tau)}$ (for $n=1,2$), which is related to baryon-to-photon ratio $\eta_{n}(t) = \frac{\rho_{n}(t)}{n_\gamma(t)}$. The inset shows that $\tilde{\eta}_{n}(\tau)$ remains at stable constant values in the long-time limit, $\tau \to \infty$.}
\label{fig-eta-til-12}
\end{figure}

Since the behaviour of $\tilde{n}_\gamma(\tau)$ is already known from Eq.~\ref{eq-ngam}, we need the behaviour of $\tilde{\rho}_{n}(\tau)$ to obtain the above ratio $\tilde{\eta}_{n}(\tau)$ so that we can analyze the baryon-to-photon ratio $\eta_{n}(t)$ from Eq.~\ref{eq-eta}. 

To obtain the functional behaviour of $\tilde{\rho}_{n}(\tau)$, we shall numerically integrate the dynamical equation for $\tilde{\rho}_{n}(\tau)$ given by \ref{eq-til-rho1}, along with the coupled equations \ref{eq-x} and  \ref{eq-y}, supplemented with \ref{eq-rhon1}.

We shall consider each of the cases $n=1,2,3$ separately for the purpose of numerical integration.

\begin{figure}
\centering
\includegraphics[width=0.70\textwidth]{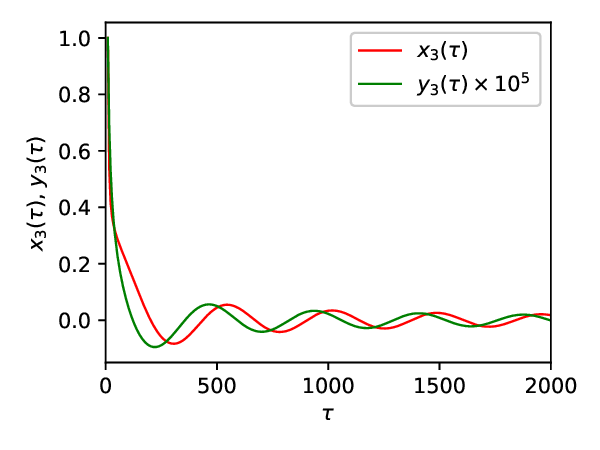}
  \caption{Temporal evolution of the dimensionless real and imaginary parts, $x_{3}(\tau)$ and $y_{3}(\tau)$, of the complex scalar field, obtained from numerically solving the coupled differential equations \ref{eq-x} and \ref{eq-y} for $n=3$.}
 \label{fig-xy-3}
\end{figure}

\section{Numerical Analysis and Results}
\label{sec-numerical}

In this section, we present a detailed numerical study of the system by solving the coupled evolution equations for the field components and the associated Noether charge density. Specifically, the dimensionless variables $x_n(\tau)$ and $y_n(\tau)$ are obtained through numerical integration of Eqs.~\ref{eq-x} and \ref{eq-y}, while the evolution of the dimensionless charge density $\tilde{\rho}_n(\tau)$ is determined self-consistently using Eq.~\ref{eq-til-rho1}, supplemented by Eq.~\ref{eq-rhon1}. The integration is performed with initial conditions corresponding to a baryon-symmetric state, ensuring $\tilde{\rho}_n(0)=0$, and the subsequent evolution captures the generation of asymmetry driven by nonlinear interactions in an expanding background. Although the governing equations are identical for all values of $n$, the resulting dynamics exhibit both shared characteristics and distinct features depending on $n$. For clarity, we therefore organize the discussion into separate subsections corresponding to different values of $n$, allowing a more transparent comparison of their qualitative and quantitative behavior.

\begin{figure}
\centering
\includegraphics[width=0.70\textwidth]{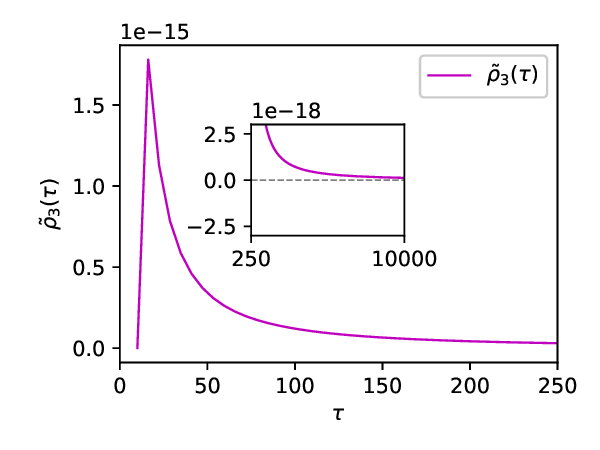}
  \caption{Temporal evolution of the dimensionless Noether charge density $\tilde{\rho}_{3}(\tau)$, obtained from simultaneous numerical solution of Eq.~\ref{eq-til-rho1}, \ref{eq-x}, and \ref{eq-y}, supplemented with Eq.~\ref{eq-rhon1}. The inset shows the long-time behaviour.} 
  \label{fig-r-til-3}
\end{figure}

\subsection{Cases $n=1,2$}
We first numerically analyze the cases $n=1,2$ because they give rise to similar features distinct from the case $n=3$. 

Figure~\ref{fig-xy-12} shows the time evolution of the dimensionless real and imaginary components, $x_{n}(\tau)$ and $y_{n}(\tau)$, of the complex scalar field, obtained from numerical integration of the coupled differential equations \ref{eq-x} and \ref{eq-y} for $n=1,2$. The plots clearly show damped oscillations arising from the nonlinear source terms on the right-hand sides of those equations in the presence of Hubble expansion. Even though the system begins in a baryon-symmetric state, the nonlinear coupling causes the two components to gradually acquire a phase difference.

Figure~\ref{fig-r-til-12} displays the time evolution of the dimensionless Noether charge density $\tilde{\rho}_{n}(\tau)$ for $n=1,2$, obtained from simultaneous numerical integration of Eq.~\ref{eq-til-rho1}, \ref{eq-x}, and \ref{eq-y}, supplemented with Eq.~\ref{eq-rhon1}. The profiles exhibit an initial sharp rise, indicating rapid generation of Noether charge density, followed by a prompt decay whose rate gradually decreases with time, and asymptotically approaches zero, as shown in the inset.

\begin{figure}
\centering
\includegraphics[width=0.70\textwidth]{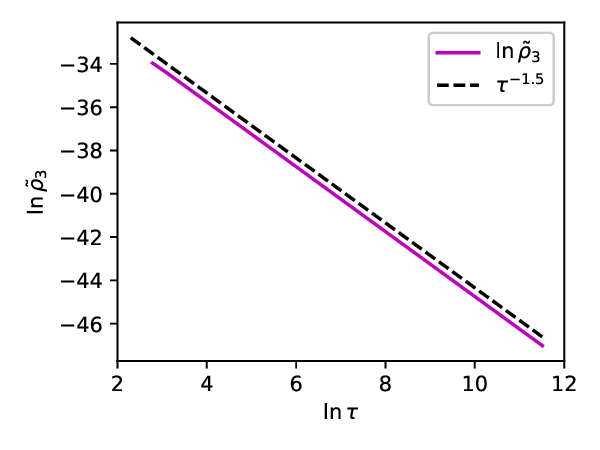}
  \caption{Logarithmic plot for the dimensionless Noether charge density, $\ln\tilde{\rho}_{3}(\tau)$ versus $\ln\tau$. The dashed reference line has a slope of $-\frac{3}{2}$.}
  \label{fig-log-r-til-3}
\end{figure}

Figure~\ref{fig-log-r-til-12} shows logarithmic plots for the dimensionless Noether charge density, $\ln\tilde{\rho}_{n}(\tau)$ versus $\ln\tau$, for $n=1,2$. The long-time behavior asymptotically becomes parallel to the dashed reference line of slope $-\frac{3}{2}$, indicating the power law  $\tilde{\rho}_{n}(\tau)  \sim \tau^{-3/2}$ in the long time limit $\tau \to \infty$.

We find numerically that as $ \tau \to \infty $, $ \tilde{\rho}_{1,2}(\tau) $ stabilize to the behaviours
\begin{equation}
\tilde{\rho}_{1}(\tau) = 2.2319 \times 10^{-4}\, \tau^{-3/2} \quad \text{and} \quad \tilde{\rho}_{2}(\tau) = 3.7583 \times 10^{-4}\, \tau^{-3/2}.
\label{eq-result-12}
\end{equation}

Figure~\ref{fig-eta-til-12} illustrates temporal evolution of $\tilde{\eta}_{n}(\tau) = \frac{\tilde{\rho}_{n}(\tau)}{\tilde n_\gamma(\tau)}$ (for $n=1,2$), which is related to  baryon-to-photon ratio $\eta_{n}(t) = \frac{\rho_{n}(t)}{n_\gamma(t)}$ via Eq.~\ref{eq-eta}. The behavior of $\tilde{\eta}_{n}(\tau)$ reveals a swift generation of charge asymmetry at early times.
After this quick rise, $\tilde{\eta}_{1}(\tau)$ exhibits a continuous decay, subsequently settling into a constant value at late times. However, $\tilde{\eta}_{2}(\tau)$ settles into a constant value after the quick rise. The inset shows that both $\tilde{\eta}_{1}(\tau)$ and $\tilde{\eta}_{2}(\tau)$ stabilize to distinct constant values in the asymptotic long-time limit, $\tau \to \infty$, indicating the approach to finite baryon charge asymmetry, although the system is initialized with a baryon-symmetric state, $\tilde{\rho}_{n}(0)=0$.

Using Eqs.~\ref{eq-til-eta}, the numerical results \ref{eq-result-12} indicate 
\begin{equation}
\tilde{\eta}_{1}(\infty) = 2.2319 \times 10^{-4} \quad \text{and} \quad  \tilde{\eta}_{2}(\infty) = 3.7583 \times 10^{-4}.
\label{eq-eta12}
\end{equation}

\begin{figure}
\centering
\includegraphics[width=0.70\textwidth]{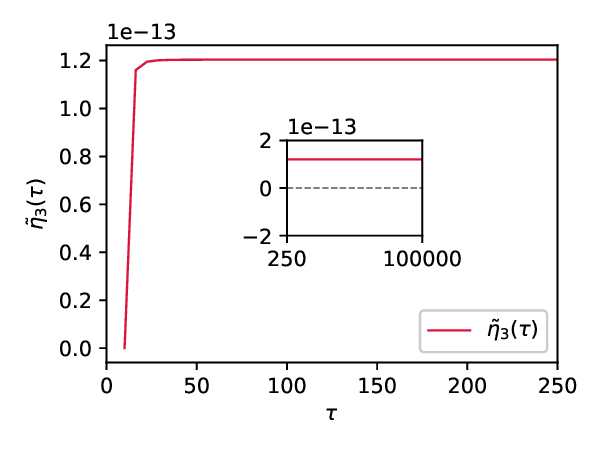}
\caption{Temporal evolution of $\tilde{\eta}_{3}(\tau) = \frac{\tilde{\rho}_{3}(\tau)}{\tilde n_\gamma(\tau)}$, which is related to baryon-to-photon ratio $\eta_{3}(t) = \frac{\rho_{3}(t)}{n_\gamma(t)}$. The inset shows that $\tilde{\eta}_{3}(\tau)$ remains at a stable constant value in the long-time limit, $\tau \to \infty$.}
\label{fig-eta-til-3}
\end{figure}

From Eq.~\ref{eq-eta}, we may write
\begin{equation}
\label{eq-eta-inf-1}
\eta_{1}(\infty) 
= \frac{1}{\lambda^{\frac{1}{2}}}\frac{\pi^{7/2}}{2\zeta(3)}\left(\frac{2g_*}{45}\right)^{3/4} \left(\frac{M}{M_{P}}\right)^{1/2} \,\tilde{\eta}_{1}(\infty)
\end{equation}
\begin{equation}
\label{eq-eta-inf-2}
\eta_{2}(\infty) = \frac{1}{\lambda^{1/3}}\frac{\pi^{7/2}}{2\zeta(3)}\left(\frac{2g_*}{45}\right)^{3/4} \left(\frac{M}{M_{P}}\right)^{1/6}\,\tilde{\eta}_{2}(\infty)
\end{equation}
which, using Eq.~\ref{eq-eta12} translate to
\begin{equation}
\label{eq-eta-inf-11}
\eta_{1} = \frac{1.64009\times 10^{-2}}{\lambda^{\frac{1}{2}}} \left(\frac{M}{M_{P}}\right)^{\frac{1}{2}}
\end{equation}
\begin{equation}
\label{eq-eta-inf-21}
\eta_{2} = \frac{2.76175\times 10^{-2}}{\lambda^{\frac{1}{3}}} \left(\frac{M}{M_{P}}\right)^{\frac{1}{6}},
\end{equation}
with $g_*=106.75$, the effective number of relativistic degrees of freedom in the Standard Model of particle physics.

The above expressions \ref{eq-eta-inf-11} and \ref{eq-eta-inf-21} indicate that the baryon-to-photon ratio $\eta_1$ is determined by the value of $M/\lambda$, while $\eta_2$ is determined by the value of $M/\lambda^2$. We may recall that $M$ is the mass of the complex scalar field and $\lambda$ is the strength of the nonlinear self-interaction potential.

%--------------------------------------------------------- 
\begin{table}[h!]
\centering
\caption{\centering \small Required mass scales $M$ and coupling constant $\lambda$ for observational constraint on $\eta_1$.}
\begin{tabular}{c|c|c|c}
\hline
$\lambda$ & $M$ (GeV) & $M/\lambda$ (GeV) & $\eta_1$  \\
\hline
$1$        & $3.35 \times 10^{3}$   & $3.35 \times 10^{3}$ & $6.1 \times 10^{-10}$ \\
$10^{-1}$  & $335$                  & $3.35 \times 10^{3}$ & $6.1 \times 10^{-10}$ \\
$10^{-2}$  & $33.5$                 & $3.35 \times 10^{3}$ & $6.1 \times 10^{-10}$ \\
$10^{-3}$  & $3.35$                 & $3.35 \times 10^{3}$ & $6.1 \times 10^{-10}$ \\
$10^{-6}$  & $3.35 \times 10^{-3}$  & $3.35 \times 10^{3}$ & $6.1 \times 10^{-10}$ \\
$10^{-9}$  & $3.35 \times 10^{-6}$  & $3.35 \times 10^{3}$ & $6.1 \times 10^{-10}$ \\
$10^{-12}$ & $3.35 \times 10^{-9}$  & $3.35 \times 10^{3}$ & $6.1 \times 10^{-10}$ \\
$10^{-15}$ & $3.35 \times 10^{-12}$ & $3.35 \times 10^{3}$ & $6.1 \times 10^{-10}$ \\
\hline
\end{tabular}
\label{tab-1}
\end{table}
%-------------------------------------------------------------------

Thus, the observational value $\eta_1 = 6.1 \times 10^{-10}$ constrains the value $M/\lambda = 3.35 \times 10^{3}$ GeV for the model $n=1$, while $\eta_2 = 6.1 \times  10^{-10}$ constrains the value $M/\lambda^2 = 2.84 \times 10^{-28}$ GeV for the model $n=2$.

Table \ref{tab-1} displays a comparison of the mass scale $M$ obtained from the observational constraint on $\eta_1$ for different values of the coupling constant $\lambda$. For $\lambda=1$, the mass scale is in the TeV range whereas for $\lambda=10^{-1}$, the mass scale is a bit higher than the SM scale. As it reduces to $\lambda=10^{-3}$, the mass scale is in the GeV range. However, for increasingly lower values of the coupling, the mass scale decreases to very low values.   

Table \ref{tab-2} displays a comparison of the mass scale $M$ obtained from the observational constraint on $\eta_2$ for different values of the coupling constant $\lambda$. It is obvious from the table that for any value of $\lambda$, the mass scale is unrealistically very low. 

We thus find that the class of potentials with $n=1$ yields a viable parameter space over a wide range of scalar masses, whereas the other class with $n=2$ requires unrealistically suppressed mass scales.

%----------------------------------------------------------------------
\begin{table}[h!]
\centering
\caption{\centering \small Required mass scales $M$ and coupling constant $\lambda$ for observational constraint on $\eta_2$.}
\begin{tabular}{c|c|c|c}
\hline
$\lambda$ & $M$ (GeV) & $M/\lambda^{2}$ (GeV) & $\eta_2$  \\
\hline
$1$        & $2.84 \times 10^{-28}$ & $2.84 \times 10^{-28}$  & $6.1 \times 10^{-10}$ \\
$10^{-1}$   & $2.84 \times 10^{-30}$ & $2.84 \times 10^{-28}$ & $6.1 \times 10^{-10}$ \\
$10^{-2}$   & $2.84 \times 10^{-32}$ & $2.84 \times 10^{-28}$ & $6.1 \times 10^{-10}$ \\
$10^{-3}$   & $2.84 \times 10^{-34}$ & $2.84 \times 10^{-28}$ & $6.1 \times 10^{-10}$ \\
$10^{-6}$   & $2.84 \times 10^{-40}$ & $2.84 \times 10^{-28}$ & $6.1 \times 10^{-10}$\\
$10^{-9}$   & $2.84 \times 10^{-46}$ & $2.84 \times 10^{-28}$ & $6.1 \times 10^{-10}$\\
$10^{-12}$  & $2.84 \times 10^{-52}$ & $2.84 \times 10^{-28}$ & $6.1 \times 10^{-10}$\\
$10^{-15}$  & $2.84 \times 10^{-58}$ & $2.84 \times 10^{-28}$ & $6.1 \times 10^{-10}$\\
\hline
\end{tabular}
\label{tab-2}
\end{table}
%--------------------------------------------------------------------

\subsection{Case $n=3$}
We now turn to the case $n=3$, which, while governed by the same set of equations, exhibits behavior that is best discussed separately for clarity.  

The evolution of the field components $x_{3}(\tau)$ and $y_{3}(\tau)$, obtained by numerically solving Eqs.~\ref{eq-x} and \ref{eq-y}, is presented in Figure~\ref{fig-xy-3}. As in the previous cases, the dynamics is characterized by damped oscillations driven by the nonlinear source terms, in an expanding background. The buildup of a relative phase between the real and imaginary components is evident as the system evolves from its initially baryon-symmetric configuration.

The corresponding dimensionless Noether charge density $\tilde{\rho}_{3}(\tau)$ is shown in Figure~\ref{fig-r-til-3}. This quantity is computed through the coupled evolution of Eqs.~\ref{eq-til-rho1}, \ref{eq-x}, and \ref{eq-y}, together with Eq.~\ref{eq-rhon1}. The profile indicates a rapid initial growth of the Noether charge density, followed by a decay phase whose rate progressively slows down, eventually approaching zero at late times, as highlighted in the inset.

Further insight into the late-time behavior is provided by the logarithmic representation in Figure~\ref{fig-log-r-til-3}, where $\ln \tilde{\rho}_{3}(\tau)$ is plotted against $\ln \tau$. The alignment with a reference line of slope $-\frac{3}{2}$ confirms the power-law decay $\tilde{\rho}_{3}(\tau) \sim \tau^{-3/2}$ in the limit $\tau \to \infty$.

Our numerical analysis shows that $ \tilde{\rho}_{3}(\tau) $ follows a well-defined asymptotic behavior
\begin{equation}
\tilde{\rho}_{3}(\tau) = 1.2038 \times 10^{-13}\, \tau^{-3/2}.
\label{eq-result-3}
\end{equation}

Figure~\ref{fig-eta-til-3} depicts the evolution of the ratio $\tilde{\eta}_{3}(\tau) = \frac{\tilde{\rho}_{3}(\tau)}{\tilde n_\gamma(\tau)}$, related to the baryon-to-photon ratio via Eq.~\ref{eq-eta}. The generation of asymmetry takes place quickly at the early time, after which $\tilde{\eta}_{3}(\tau)$ levels off to a constant value. As seen in the inset, it remains a constant at late times, signaling the emergence of a finite baryon asymmetry despite the symmetric initial condition $\tilde{\rho}_{3}(0)=0$.  

Using Eqs.~\ref{eq-til-eta}, the numerical result \ref{eq-result-3} indicates 
\begin{equation}
\tilde{\eta}_{3}(\infty) = 1.2038 \times 10^{-13}.
\label{eq-eta3}
\end{equation}
From Eq.~\ref{eq-eta}, we may write
\begin{equation}
\label{eq-eta-inf-3}
\eta_{3}(\infty) =  \frac{1}{\lambda^{1/4}}\frac{\pi^{7/2}}{2\zeta(3)}\left(\frac{2g_*}{45}\right)^{3/4} \,\tilde{\eta}_{3}(\infty).
\end{equation}
which, using Eq.~\ref{eq-eta3}, translates to
\begin{equation}
\label{eq-eta-inf-31}
\eta_{3} = \frac{8.846007\times 10^{-12}}{\lambda^{\frac{1}{4}}}. 
\end{equation}

Thus, the baryon-to-photon ratio, in this case of interaction potential, is independent of the mass scale $M$ of the scalar field and depends only on the coupling parameter $\lambda$. This feature is particularly a strong point with this model since it requires only one parameter $\lambda$ although the other parameter $M$ is present in the theory.

Thus, the observational value $\eta_3= 6.1 \times 10^{-10}$ constrains the coupling parameter to 
\begin{equation}
\lambda = 4.423 \times 10^{-8}.
\end{equation}

This value is intriguingly comparable to typical constraint on the coupling parameter of the $\phi^{4}$ theory of inflation.

\section{Discussion and Conclusion}
\label{sec-disc}

In this work, we have presented a non-supersymmetric framework for baryogenesis driven by the nonlinear dynamics of a complex scalar field with explicit $U(1)$-breaking self-interaction potentials in a radiation-dominated Universe. Specifically, we took three different forms of the self-interaction potential to analyze the mechanism of baryogenesis. The resulting analysis exhibits a variety of results, depending on the specific form of the non-linear interaction, highlighting the sensitivity of the asymmetry generation mechanism to microscopic form of the interaction.

In each case, detailed numerical investigation of the coupled evolution of the real and imaginary components of the scalar field leads to the generation of Noether charge asymmetry, driven by non-linear source terms originating from the $U(1)$-breaking potential, even though starting from symmetric initial condition. The emergence of different dynamical trajectories for the field components provides a purely dynamical origin for Noether charge generation in the radiation-dominated Universe.

Our results demonstrate that the specific forms of the nonlinear interaction, together with Hubble expansion, play a crucial role in generating different trajectories of the field components, which in turn leads to the dynamical generation of Noether charge. At late times, the Noether charge density is found to behave like $\rho_n(t) \propto t^{-3/2}$, which is identical to the scaling of the photon number density. Consequently, the ratio $\eta(t) = \frac{\rho_n(t)}{n_\gamma(t)}$ approaches a constant at late times, indicating dynamical freeze-in of the baryon-to-photon ratio.

Under the assumption of efficient transfer of the generated asymmetry to the Standard Model baryonic sector and negligible late-time washout effects, the generated asymmetry can be identified with the observed baryon asymmetry of the Universe.

While the qualitative features of the evolution remains similar across different cases of interaction potential, our analysis reveals that the specific form of the interaction potential influences the quantitative aspects of the dynamics, including the rate of charge generation and the final saturation value of the asymmetry.

In the first case of interaction potential, the observed value of baryon-to-photon ratio constrains the ratio of scalar mass to the coupling strength to $\sim 10^{3}$ GeV, allowing for a wide range of viable mass scales from the TeV regime down to much smaller values.

On the other hand, in the second case of interaction potential, the observational constraint requires the ratio of the scalar mass to the square of the coupling to be of order $\sim 10^{-28}$ GeV, implying an extremely suppressed unrealistic mass scale and rendering the model phenomenologically unviable.

A key result is given by the third model of interaction potential. The baryon-to-photon ratio becomes independent of the scalar mass and depends only on the coupling parameter $\lambda$. This feature is particularly a strong point with this model since it effectively reduces the parameter space and enhances predictivity. Imposing the observational constraint yields $\lambda \sim 10^{-8}$, a value intriguingly comparable to typical constraint on the coupling parameter of the $\phi^{4}$ theory of inflation.

Consequently, the third model is flexible across a broad range of scalar field masses. This parametric dependence suggests that the mechanism may operate naturally across multiple energy scales, potentially connecting early-Universe dynamics with high-energy scalar sectors beyond the Standard Model.

Thus this study demonstrates that nonlinear scalar field dynamics alone, even in the absence of supersymmetry, can provide a viable mechanism for baryogenesis. The robustness and flexibility of the mechanism motivate further exploration of non-supersymmetric early-Universe models in which cosmic asymmetries arise from purely dynamical effects.

\section{Acknowledgements}

Surendra Kumar Gour gratefully acknowledges financial support through a Research Fellowship from the Ministry of Education, Government of India. The Authors would like to thank the Indian Institute of Technology Guwahati for providing access to computing facilities.

%\bibliographystyle{unsrturl}
%\bibliography{bb}
%\end{multicols}

\end{document}